\documentclass[manuscript]{aastex61}

\newcommand{\alf}{Alfv\'en }
\newcommand{\ALF}{ALFV\'EN }

\newcommand{\be}{\begin{equation}}
\newcommand{\ee}{\end{equation}}
\newcommand{\bea}{\begin{eqnarray}}
\newcommand{\eea}{\end{eqnarray}}
\newcommand{\bean}{\begin{eqnarray*}}
\newcommand{\eean}{\end{eqnarray*}}

\submitjournal{ApJ}

\shorttitle{Threaded-Field-Line Model}
\shortauthors{Sokolov et al.}
\begin{document}

\title{Threaded-Field-Line Model for the Low Solar Corona Powered by
  the \ALF Wave Turbulence}

\correspondingauthor{Igor Sokolov}
\email{igorsok@umich.edu}

\author{Igor V. Sokolov}
\affil{Climate and Space Sciences and Engineering, University of Michigan,
2455 Hayward St, Ann Arbor, MI 48109, USA}

\author{ Bart van der Holst}
\affiliation{Climate and Space Sciences and Engineering, University of Michigan,
2455 Hayward St, Ann Arbor, MI 48109, USA}

\author{Ward B. Manchester}
\affiliation{Climate and Space Sciences and Engineering, University of Michigan,
2455 Hayward St, Ann Arbor, MI 48109, USA}

\author{Doga Can Su Ozturk}
\affiliation{Climate and Space Sciences and Engineering, University of Michigan,
2455 Hayward St, Ann Arbor, MI 48109, USA}

\author{Judit Szente}
\affiliation{Climate and Space Sciences and Engineering, University of Michigan,
2455 Hayward St, Ann Arbor, MI 48109, USA}

\author{Aleksandre Taktakishvili}
\affiliation{Community Coordinated Modeling Center, NASA Goddard Space Flight 
Center, Greenbelt, MD 20771, USA}

\author{G\'abor T\'oth}
\affiliation{Climate and Space Sciences and Engineering, University of Michigan,
2455 Hayward St, Ann Arbor, MI 48109, USA}

\author{Meng Jin}
\affiliation{Lockheed Martin Solar and Astrophysics Lab, Palo Alto, CA 94304, USA}

\author{Tamas I. Gombosi}
\affiliation{Climate and Space Sciences and Engineering, University of Michigan,
2455 Hayward St, Ann Arbor, MI 48109, USA}



\begin{abstract}

We present an updated global model of the solar corona,
including the transition region. We simulate the realistic tree-dimensional (3D)
magnetic field using the data from the photospheric
magnetic field measurements and assume the magnetohydrodynamic (MHD) Alfv\'en wave turbulence
and its non-linear dissipation to be the only 
source for heating the coronal plasma and driving the solar wind. In closed field regions the dissipation efficiency in a balanced turbulence is enhanced.  In the coronal holes we account for a reflection of the outward propagating waves, which is accompanied by generation of weaker counter-propagating waves. The non-linear cascade rate degrades in strongly imbalanced turbulence, thus resulting in colder coronal holes. 

The distinctive feature of the presented model is the description of the low corona as
almost-steady-state low-beta plasma motion and heat flux transfer along the magnetic field
lines. We 
trace the magnetic field lines through each grid point of the lower 
boundary of the global corona model, chosen at some 
heliocentric distance,  $R=R_{b}\sim1.1\ R_\odot$ well above the transition region. One can readily solve the 
plasma parameters along the magnetic field line from 1D equations for the 
plasma motion and heat transport together with the Alfv\'en wave propagation, which adequately describe physics within the heliocentric distances range, $R_{\odot}<R<R_{b}$, in the low solar corona. By
interfacing this threaded-field-lines model with the full MHD global corona model at 
$r=R_{b}$, we find the global solution and achieve a faster-than-real-time performance of the model on $\sim200$ cores.

\end{abstract}



\section{Introduction}
Observations from Hinode  and Solar Dynamics Observatory  (SDO) (\cite{dupo08} and \cite{mcintosh11}) raised the estimate for the  \alf wave energy in the solar corona (SC). About $10\div20\%$ of this outward propagating energy is adequate to heat the SC and accelerate the solar wind in inner heliosphere (IH). Therefore, several three-dimensional (3-D) solar wind (\cite{usma00},  \cite{suzu05}, \cite{verd10},  \cite{osman11}, {\color{cyan}{\it Lionello et al.} (2014a, 2014b)}) and coronal heating  (\cite{tu97}, \cite{hu00}, \cite{ dmit02}, \cite{Habb03} and \cite{cran10}) models that included, or were  exclusively driven by, \alf wave turbulence became increasingly popular and paved the road for the development of even more advanced \alf wave driven models.

Although popular, this physics-based approach to modeling the solar environment is not the only way to model the solar corona and the solar wind. Semi-empirical descriptions of the solar wind, like the widely used Wang-Sheeley-Arge (WSA) model \cite[] {arge00} is also attractive because of their simplicity and ability to predict the solar wind speed in the IH. In addition, the WSA formulae can be easily incorporated  into global 3-D models for the SC and IH  \cite[see ][]{cohen07} via a varying polytropic index distribution as proposed by \cite{roussev03b}.  Similarly, instead of the \alf wave turbulence dissipation mechanism to heat the corona, one can use  well established models with semi-empirical heating functions, such as those presented by {\color{cyan}{\it Lionello et al.} (2001, 2009)}, \cite{rile06}, \cite{tito08} and \cite{downs10}. This method leads to reasonably good agreements with observations in EUV, X-rays and white light. The agreement looks particularly impressive for the PSI predictions about the solar eclipse image.

An important limitation of the semi-empirical models is that they depend on free parameters that need to be determined for various solar conditions. This fact makes it complicated to use them in an integrated modeling approach describing the SC and IH system with very few free parameters. In the presented research, the  \alf wave turbulence is treated as the only energy source to heat the SC and to power and accelerate the solar wind.
 
From the model for the quiet-time SC and IH the {\it ad hoc} elements were eliminated  by \cite{sok13}. In the Alfv\'en-Wave-driven Solar atmosphere Model (AWSoM) the plasma is heated by the dissipation of the Alfv\'en wave turbulence, which, in turn, is generated by the nonlinear interaction between oppositely propagating waves \cite[]{hollw86}. Within the coronal holes there are no closed magnetic field lines, hence, there are no oppositely propagating waves. Instead, a weak reflection of the outward propagating waves locally generate sunward propagating waves as quantified by \cite{vanderholst13}. The small power in these locally generated (and almost immediately dissipated) inward propagating waves leads to a reduced turbulence dissipation rate in coronal holes, naturally resulting in the bimodal solar wind structure. Another consequence is that coronal holes look like cold black spots in the EUV and X-rays images, the closed field regions are  hot and bright, and the brightest are active regions, near which the wave reflection is particularly strong (see \cite{sok13}, \cite{oran13} and \cite{vanderholst13}). 

The described global models simulate the steady state of the solar terrestrial environments, which 
serves as a background for {\it space weather}.   
Space weather describes the dynamic state of the Earth's
magnetosphere-ionosphere system, which is driven by the solar wind and solar
ionizing radiation.  The greatest disturbances in space weather are
geomagnetic storms, the most severe of which are caused by coronal mass
ejections (CMEs) (see \cite{gosling1993}). To simulate the evolution of such ``transient'' from Sun to 1 AU requires significant computational resources.

Here we present the development of the AWSoM. A distinctive feature of the presented model is the description of the low SC as
almost-steady-state low-beta plasma motion along the magnetic field
lines, the heat fluxes also being aligned with the magnetic field.  The  Low Solar Corona model which ranges from the upper chromosphere to the heliocentric distances about 
$\sim 1.1\ R_{\odot}$ and includes the transition region at $R_\odot<R<1.03\ R_\odot$, is the heart of the global models.
 In the low SC the Alfv{\'e}n waves pass from the chromosphere to the solar corona, 
the plasma temperature increases by two orders of magnitude (from ten thousand to million K),
and this is also a place where the solar wind originates. The multi-wavelength observations (in
 EUV and X-rays) from several satellite locations (SDO, STEREO A,B) may be used to validate 
the simulation model. Therefore, any global model must account for the processes in this region.
On the other hand, for the simulation model to explain the space weather and also have a 
predictive capability, it should be capable of simulating the dynamic processes faster than 
they proceed in real time, and the low SC appears to be a bottleneck limiting the 
computational efficiency and performance. 
  
In numerical simulations of the solar corona, both for the ambient state and especially for 
dynamical processes, the greatest number of computational resources is spent for maintaining 
the numerical solution in the low SC and in the transition region, where the 
temperature gradients are sharp and the magnetic field topology is complicated. The degraded 
computational efficiency is caused by the need for the highest resolution as well as the use 
of a fully three-dimensional implicit solver for electron heat conduction. The need to find 
a numerical method, which would allow us to gain in the computational efficiency, 
motivates the research presented here.    

We benefit from the observation that although the simulations of the low SC are 
computationally intense, the physical nature of the processes involved is rather simple as 
long as the heat fluxes and slow plasma motional velocities are mostly aligned with the 
magnetic field. The Alfv\'en wave turbulence, is characterized by the wave Poynting flux, 
which is also aligned with the magnetic field. Therefore, the plasma state at any point within 
the low SC is controlled by the plasma, particle, and Alfv\'en wave transport along 
the magnetic field line, which passes through this point.  This physical property is typical 
for a variety of magnetized plasmas in different astrophysical and laboratory environments and 
may be used as the base of a new numerical method, which solves the state of plasma in each 
grid point in the computational domain depth in the following way: (1) by passing the magnetic 
field line ('thread') through this point and connecting it with the domain boundaries (e.g., 
with chromosphere and with the global solar corona domain, once the method is applied to the 
low SC) and (2) by solving a set of one-dimensional transport equations to relate 
the grid point value to the boundary conditions.

We trace the magnetic field lines through all grid points of the lower boundary of the global coronal model chosen at some heliocentric distance $R=R_b\sim1.1\ R_\odot$ well above the transition region. One can readily solve the plasma parameters along the magnetic field line from effectively 1D equations for the plasma motion and heat transfer together with the Alfv\'en wave
propagation, which adequately describe physics within the heliocentric distance range, 
$R_\odot<R<R_b$, i.e. in the low solar corona.  By
interfacing this Threaded-Field-Line Model (TFLM) for the low corona
with full MHD global corona model at $R=R_b$ we find the global solution and achieve faster-than-realtime performance of  the model with moderate computational resources. Due to the latter feature we call the newly developed model AWSoM-R (AWSoM-Realtime).
\section{MHD Equations for the Transition Region, Solar Corona and Inner Heliosphere}
\subsection{Full 3D Governing Equations of the Global Model}
The global model within the range of heliocentric distances, $R_b<R<1\div3$ AU, $R_b\sim1.1\ R_\odot$ employs the standard MHD equations (non-specified
denotations are as usually): 
\begin{equation}\label{eq:cont}
\frac{\partial\rho}{\partial t}+\nabla\cdot(\rho{\bf u})=0,
\end{equation}
\begin{equation}\label{eq:induction}
\frac{\partial{\bf B}}{\partial t}+\nabla\cdot\left({\bf u}{\bf
  B}-{\bf B}{\bf u}\right)=0,
\end{equation}
\begin{equation}\label{eq:momentum}
\frac{\partial(\rho{\bf u})}{\partial t}+\nabla\cdot\left(\rho{\bf
  u}{\bf u}-\frac{{\bf B}{\bf
    B}}{\mu_0}\right)+\nabla\left(P_i+P_e+\frac{B^2}{2\mu_0}+P_A\right)=-\frac{GM_\sun\rho{\bf
    R}}{R^3},
\end{equation}
(herewith, $B=|{\bf B}|$), with the full energy equations applied separately to ions 
\begin{eqnarray}\label{eq:energy}
\frac{\partial}{\partial
  t}\left(\frac{P_i}{\gamma-1}+\frac{\rho u^2}2+\frac{{\bf B}^2}{2\mu_0}\right)+\nabla\cdot\left[\left(\frac{\rho u^2}2+\frac{\gamma P_i}{\gamma-1}+\frac{B^2}{\mu_0}\right){\bf
  u}-\frac{{\bf B}({\bf u}\cdot{\bf B})}{\mu_0}\right]=\nonumber\\
= -{\bf u}\cdot\nabla\left(P_e+P_A\right)+
\frac{N_eN_ik_B}{\gamma-1}\left(\frac{\nu_{ei}}{N_i}\right)\left(T_e-T_i\right)+f_p\left(\Gamma_-w_-+\Gamma_+w_+\right)-\frac{GM_\sun\rho{\bf
    R}\cdot{\bf u}}{R^3},
\end{eqnarray}
and to electrons:
\begin{eqnarray}\label{eq:electron}
\frac{\partial\left(\frac{P_e}{\gamma-1}\right)}{\partial
  t}&+&\nabla\cdot\left(\frac{P_e}{\gamma-1}{\bf
  u}\right)+P_e\nabla\cdot{\bf u}=\nonumber\\
&=&\nabla\cdot\left(\kappa\cdot\nabla 
T_e\right)+\frac{N_eN_ik_B}{\gamma-1}\left(\frac{\nu_{ei}}{N_i}\right)\left(T_i-T_e\right)-Q_{\rm rad}+(1-f_p)\left(\Gamma_-w_-+\Gamma_+w_+\right),
\end{eqnarray}
where, for a hydrogen plasma, $N_e=N_i=\rho/m_p$, $m_p$ being the proton mass. In addition to the standard effects, the above equations account for the radiation energy loss from an optically thin plasma, $Q_{\rm rad}=N_eN_i\Lambda_R(T_e)$, a
possible difference in the electron and ion temperatures, $T_{e,i}$, the electron
heat conduction parallel to the magnetic field lines equals:
\begin{equation}
\kappa={\bf b}{\bf b}\kappa_\|,\quad \kappa_\|=3.2\frac{6\pi}{\Lambda_C}
         \sqrt{\frac{2\pi}{m_e}}\left (\frac{\varepsilon_0}{e^2}\right)^2(k_BT_e)^{5/2}k_B, \qquad {\bf b}={\bf B}/B,
\end{equation}
where $m_e$ and $e$ are the electron mass and charge correspondingly, $\Lambda_C$ being the Coulomb logarithm. Eq.~6 is applied only at lower heliocentric distances, $R<6R_\sun$. Above this region, at $R>6R_\sun$, we apply ``collisionless'' heat conduction in the manner described by \cite{vanderholst13}. 

The energy exchange between electron and ions is
parameterized via the energy exchange rate, $\frac{\nu_{ei}}{N_i}= \frac{2\sqrt{m_e}\Lambda_{C}
	(e^2/\varepsilon_0)^2
}{
         3 m_p(2\pi k_BT_e)^{3/2}}$, as this is usually
done. The Alfv\'en wave turbulence pressure, $P_A=(w_-+w_+)/2$, and
dissipation rate, $\Gamma_-w_-+\Gamma_+w_+$, are
applied in the above equations. Herewith, $w_{\pm}$  are the
energy densities for the turbulent waves propagating along the magnetic
field vector ($w_+$) or in the opposite direction ($w_-$). The turbulence energy dissipation (see Eqs.\ref{eq:energy}-\ref{eq:electron}) is split into electron and ion heating.  The ion heating fraction, $f_p$, is quantified using the physics-based model for partitioning energy between ions and electrons  (see \cite{vanderholst13} and the papers cited therein). At higher densities as in the low corona, we assume $T_e=T_i$ and
use the single-temperature energy equation for electron and ions, to improve the computational efficiency:
\begin{eqnarray}\label{eq:energytotal}
\frac{\partial}{\partial
  t}\left(\frac{P}{\gamma-1}+\frac{\rho u^2}2+\frac{{\bf B}^2}{2\mu_0}\right)+\nabla\cdot\left[\left(\frac{\rho u^2}2+\frac{\gamma P}{\gamma-1}+\frac{B^2}{\mu_0}\right){\bf
  u}-\frac{{\bf B}({\bf u}\cdot{\bf B})}{\mu_0}\right]=\nonumber\\
=-{\bf u}\cdot\nabla P_A +\nabla\cdot\left(\kappa\cdot\nabla 
T\right)-Q_{\rm rad}+\Gamma_-w_-+\Gamma_+w_+-\frac{GM_\sun\rho{\bf
    r}\cdot{\bf u}}{r^3},
\end{eqnarray}
where $P=P_e+P_i=2N_ik_BT$. We use the equation of state,
$P_{e,i}=N_{e,i}k_BT_{e,i}$ for the 
coronal plasma with the polytropic index, $\gamma=\frac53$. 
To complete the model, the  equation describing propagation, reflection and dissipation of turbulent waves has been derived in \cite{vanderholst13} following the approach as adopted in \cite{vell93} \cite{tu95}, \cite{dmit02}, \cite{chan09} and \cite{chan09a}):   
\begin{equation}\label{eq:wkbrefl}
\frac{\partial w_\pm}{\partial t}+\nabla\cdot\left[({\bf u}\pm{\bf
    V}_A)w_\pm\right]+\frac{w_\pm}2(\nabla\cdot{\bf u})=
\mp{\cal R}\sqrt{w_-w_+}
-\Gamma_\pm w_\pm
\end{equation}
where $\Gamma_\pm=\frac2{L_\perp}\sqrt{\frac{w_\mp}\rho}$ (note that the definition of $L_\perp$ and, accordingly the expression for $\Gamma_\pm$ used both here and in \cite{vanderholst13} are by a factor of 2 different from those used in \cite{sok13}).  The reflection coefficient has been constructed as follows:
\begin{eqnarray}{\cal R}=\min\left[\sqrt{\left({\bf b}\cdot[\nabla\times{\bf
    u}]\right)^2+\left[({\bf V}_A\cdot\nabla)\log
    V_A\right]^2},\max(\Gamma_\pm)\right]\times\nonumber\\
\times\left[\max\left(1-\frac{I_{\rm max}}{\sqrt{{w_+}/{w_-}}},0\right)-\max\left(1-\frac{I_{\rm max}}{\sqrt{{w_-}/{w_+}}},0\right)\right],
\label{eq:reflection}
\end{eqnarray} 
where $I_{\rm max}=2$ is the maximum "imbalance degree". If the "plus" wave strongly dominates, so that $\sqrt{w_+/w_-}>I_{\rm max}$, the multiplier in the second line tends to +1, in the opposite limiting case of the dominant "minus" wave, it tends to -1. In both these cases the reflection reduces the dominant wave and amplifies the minor one. Otherwise, if the both amplitude ratios do not exceed $I_{\rm max}$, the turbulence is treated as "balanced" and the reflection coefficient turns to zero. The reflection model used by \cite[]{matt99} was similar to ours.   

An important distinction is that we don't introduce the incompressible-to-compressible mode conversion term proportional to 
${\bf u}\cdot\nabla\log V_A$ into our model, although it is sometimes  accounted for by other authors. The reason for this omission is that this term would break the energy conservation in the model, because it describes the conversion to the compressible MHD turbulence, which is not included (for more detail see \cite{vanderholst13}).  

The boundary condition for the Poynting flux  at the top of chromosphere, $S_A$ is given by
$\frac{(S_A)_{R_\odot}}{B_{R_\odot}}={\rm const}=\left\{\frac{ S_A}{B}\right\}$.  Herewith, we denote with braces the constant parameters of the model, equal to a product or ratio of physical variables. 
The
estimate of the {\it constant} Pointing-flux-to-field ratio at the solar surface may be found in \cite{sok13}, \cite{oran13} and \cite{vanderholst13}:
$\left\{\frac{ S_A}{B}\right\}\approx1.1\cdot10^6\frac{\rm W}{\rm m^2T},$
where the boundary condition for the wave energy density should be
applied to the outgoing wave only.  The estimate is very close to that which follows from
\cite{Pevtsov}, \cite{,suzu06}, \cite{abbe07}, \cite{downs10} and \cite{cran10}.  
To close the model we chose, following
\cite{hollw86}, the scaling law for the transverse correlation length:
\begin{equation}\label{eq:LPerp}
L_\perp\propto B^{-1/2},\qquad 
100\,{\rm km}\ {\rm T}^{1/2}\le \{L_\perp\sqrt{B}\}\le
300\,{\rm km}\ {\rm T}^{1/2}.
\end{equation}
\subsubsection{Alternative 3D equations for Alfv\'en Wave Dynamics}
Eq.(\ref{eq:wkbrefl}) has a form close to the conservation law, which is well suited for solving it numerically within the framework of the global coronal model. However, both for using in the TFLM model and for analytical solution, an alternative form of this 
equation may be derived based on the substitution, $w_\pm=\{S_A/B\}\sqrt{\mu_0\rho}a^2_\pm$. Using the mass conservation law, one obtains:
\begin{equation}\label{eq:aplusminus}
	\frac{\partial a^2_\pm}{\partial t}+\nabla\cdot({\bf u}a^2_\pm)\pm({\bf V}_A\cdot\nabla)a^2_\pm=\mp {\cal R}a_-a_+-2\sqrt{\frac{\{S_A/B\}\mu_0V_A}{\{L_\perp\sqrt{B}\}^2}}a_\mp a^2_\pm.
\end{equation}
The plasma heating function, $\Gamma_+w_++\Gamma_-w_-$, in these variables equals 
\begin{equation}\label{eq:heatingf}
\Gamma_+w_++\Gamma_-w_-=2(a_++a_-)a_+a_-\{S_A/B\}B\sqrt{\frac{\{S_A/B\}\mu_0}{\{L_\perp\sqrt{B}\}^2V_A}}.
\end{equation}
The dimensionless amplitude, $a_\pm$, of the outgoing wave at the lower boundary of the model equals unity. In Eqs.(\ref{eq:aplusminus}) the dimensionless wave amplitudes depend on the plasma dynamical profile only via the plasma velocity
as well as the Alfv\'en speed. In the inner heliosphere,  the Alfv\'en speed is negligible compared to the solar wind speed. Assuming steady state radial solar wind motion with the constant speed (i.e. independent on the heliocentric distance), the dimensionless amplitudes, mass density and the total turbulence energy density decay with the heliocentric distance as follows: $(a^2_++a^2_-)\propto 1/R^2$, $\rho\propto 1/R^2$,  and $(w_++w_-)\propto1/R^3\propto\rho^{3/2}$, the latter relationship being in agreement with the polytropic index of $3/2$, for the Alfv\'en wave turbulence. In the low SC the Alfv\'en wave turbulence dynamics is more complicated and discussed below.
\subsection{Equations of the Threaded-Field-Line Model} 
Now,  the governing AWSoM equations may be applied to simulate the transition region and Low Solar Corona domain at $R_\odot<R<R_b$.  We present both the simplified 1D model equations for this domain and the way how the model may be interfaced both to the chromosphere at $R= R_\odot$ and to the global corona model at $R=R_b$.  
\subsubsection{Magnetic field}
The realistic model for the 3D solar magnetic field includes  the boundary condition for the
coronal magnetic field taken from the full disc magnetogram incorporating the current and past  observation results. The potential magnetic field  provides the minimum of magnetic free energy for given boundary condition, therefore, in the "ambient" solution for the solar wind the magnetic field is approximately equal to the potential one in the close proximity of the Sun. Following \cite{ogin84} and \cite{tan94} (see also \cite{powell99} and \cite{gomb02}) we split the total magnetic field ${\bf B}={\bf B}_0+{\bf B}_1$, in such way that the {\it potential} ${\bf B}_0$ field dominates at  $R=R_\odot$. If the observable is the radial component of the magnetic field at the photospheric level, then the potential ${\bf B}$ field may be recovered from the observed magnetogram using the Potential Field Source Surface Method (PFSSM) had been for the first time described in \cite{alt77}. The Laplace 
equation for scalar magnetic potential is solved at $R_\odot<R<R_{SS}=2.5R_\odot$ with the given radial gradient of the potential (the observed radial field) at $R=R_\odot$ and with vanishing magnetic potential (i.e. purely radial magnetic field) at $R=R_{SS}$, using the development into a series of spherical harmonics. Accordingly, non-potential ${\bf B}_1$ field within the original split field approach (used, particularly, in \cite{sok13}, \cite{oran13} and \cite{vanderholst13}) satisfies zero boundary condition for the radial
field component, $({\bf B}_1)_R=0$, at $R=R_\odot$, as long as the observed field is fully included into the potential field. 

The distinction of the approach presented here is that we neglect non-potential ${\bf B}_1$ field in the Low Corona and assume that ${\bf B}_1\equiv0$ at $R_\odot<R<R_b$. Accordingly, the boundary condition, $({\bf B}_1)_R=0$, is accepted within the global model at $R=R_b$. In this way we benefit in easily bridging the field observed at $R=R_\odot$ to the 
model starting at $R=R_b$. Second, the lines of the potential, ${\bf B}_0$, field at $R_\odot<R<R_b$
give us the {\it threads} which allow us to bridge the boundary conditions for all other physical quantities
from the top of chromosphere to the global model boundary at $R=R_b$.
\subsubsection{Magnetic thread and the conservation laws on it.}
Now, we introduce a key concept of the Threaded-Field-Line Model (TFLM) - a thread. 
The boundary conditions for the global model are to be applied at each grid point of the global 
model boundary at $R=R_b$.  The {\it potential} magnetic field line,  "thread", starting at the grid point can be traced through the  Low Corona domain, $R_\odot<R<R_b$ toward the Sun. 
To reduce the 3D governing equations to effectively 1D equations, one can integrate Eqs.(\ref{eq:cont},\ref{eq:momentum},\ref{eq:energytotal}) over a magnetic flux tube element of a length of $ds$ bounded by two close cross sections of the flux tube, $dS_1$ and $dS_2$, and a bundle of magnetic field lines about the considered thread, which all 
pass through the contours of these cross sections. The equation, $\nabla\cdot{\bf B}=0$ gives: $BdS={\rm const}$ along the flux tube,  which allows us to relate the change in the cross-section area along the thread to the magnetic field magnitude. The conservation laws are greatly simplified due to the fact that the velocity of low-beta plasma motion is aligned with the magnetic field, ${\bf u}=u{\bf b}$, where $u$ is a scalar aligned velocity. Particularly, the continuity equation (\ref{eq:cont}) for a steady-state flow along the flux tube gives: $\rho u S={\rm const}$, so that the ratio of constant mass flux to the constant magnetic flux gives $\frac{\partial}{\partial s}\left(\frac{\rho u}{B}\right)=0$ and 
\begin{equation}\label{eq:cont1d}
\left\{\frac{\rho u}{B}\right\}=m_p\left\{\frac{N_i u}{B}\right\}={\rm const},
\end{equation} 
where $\frac\partial{\partial s}=({\bf b}\cdot\nabla)$. Herewith we denote with braces the combinations of variables, which are constant along the thread (above we did  this only for the model  parameters, $\{S_A/B\}$ and $\{L_\perp\sqrt{B}\}$ ).  As long as the velocity is not solved within the TFLM, the constant ratio in Eq.(\ref{eq:cont1d})
is found from the Global Corona Model (GCM) side: $\left(\frac{u}B\right)_{TFLM}=\lim_{R\rightarrow R_b+0}\left(({\bf B}\cdot{\bf u})/B^2\right)_{GCM}$.

In the momentum equation we neglect $u^2$ relative to the speed of sound squared, $\gamma P/\rho$ and omit ${\bf j}\times{\bf B}$ force, vanishing in the potential magnetic field (${\bf j}\propto\nabla\times{\bf B}_0=0$), which gives us the hydrostatic equilibrium equation:
 \begin{equation}\label{eq:hydrostatic}\frac{\partial P}{\partial s}=-\frac{b_RGM_{\odot}\rho}{R^2}.
 \end{equation} 
 The latter can be integrated, if desired, for the given profile of temperature giving the barometric formula, $P=P_{TR}\exp\left[\int_{R_{TR}}^R{\frac{d(GM_\odot m_p/R)}
{2k_BT}}\right]$, the values of variables on top of the transition region (TR) are discussed below. 

In Eq.(\ref{eq:energytotal}) we keep the time derivative of temperature as long as the electron heat conduction is a comparatively slow process:
\begin{eqnarray}\label{eq:energytotal1d}
 \frac{2N_ik_B}{B(\gamma-1)}\frac{\partial T}{\partial t}&+& \frac{2k_B\gamma}{(\gamma-1)}\left\{\frac{N_iu}B\right\}\frac{\partial T}{\partial s}=\nonumber\\
= \frac\partial{\partial s}\left(\frac{\kappa_\|}B\frac{\partial T}{\partial s}\right) &+& \frac{\Gamma_-w_-+\Gamma_+w_+-N_eN_i\Lambda_R(T)}B+\left\{\frac{\rho u}{B}\right\}\frac{\partial(GM_\sun/r)}{\partial s}.
\end{eqnarray}
Note, that we neglect the Alfv\'en wave pressure gradients in Eqs.(\ref{eq:hydrostatic},\ref{eq:energytotal1d}). Practically in the Low Corona this pressure is small, being proportional to a square root of high density, while the thermal pressure is proportional to the density. Theoretically, keeping this term in Eq.(\ref{eq:energytotal1d}) would be inconsistent. Comparing with the Alfv\'en wave energy deposition (see below) it involves the small ratio of the plasma speed to the Alfv\'en wave speed, and all such terms are neglected in deriving Eq.(\ref{eq:wkbrefl1d}) below.
\subsubsection{Alfv\'en wave 1D dynamics.}
The physical 
property of the Alfv\'en waves to have the energy flux aligned with the magnetic field allows us to reduce 3D differential operators in the governing equations to the advective derivatives along the magnetic field lines. In addition, the 1D governing equations, which are obtained in this way, may be further simplified for the low corona environments. Indeed, in the low SC  the plasma velocity in Eqs.(\ref{eq:aplusminus}) is negligible compared to the Alfv\'en wave speed. The steady-state solutions for $a_\pm$  may be sought for, as long as the non-stationary perturbations propagate with the large Alfv\'en wave speed across the low corona and quickly converge to an equilibrium, so that Eqs(\ref{eq:aplusminus}) once divided by $V_A$ may be written as follows:
$
\pm({\bf b}\cdot\nabla)a^2_\pm=\mp \frac{{\cal R}}{V_A}a_-a_+-2\sqrt{\frac{\{S_A/B\}\mu_0}{\{L_\perp\sqrt{B}\}^2V_A}}a_\mp a^2_\pm
$.  This equation may be further simplified by substituting  
$d\xi=ds\sqrt{\frac{\{S_A/B\}\mu_0}{\{L_\perp\sqrt{B}\}^2V_A}}$:
\begin{equation}\label{eq:wkbrefl1d}
\pm\frac{da_\pm}{d\xi}=\mp\frac{ds}{d\xi}\frac{\cal R}{2V_A}a_\mp-a_-a_+.
\end{equation}
As long as the plasma speed is small relative to the Alfv\'en speed, the velocity curl in the expression for the reflection coefficient is negligible compared with the contribution from the Alfv\'en speed gradient, hence:   
\begin{equation}\label{eq:reflection1}\frac{ds}{d\xi}\frac{\cal R}{2V_A}=\min
\left( \frac{|d\log V_A/d\xi|}{2a_{\max}} ,1\right)
\left( \max\left( a_{\max}-2a_-,0 \right) -\max\left(a_{\max}-2a_+,0\right) \right),\end{equation}
$a_{\max}=\max(a_-,a_+)$. The formulation of the
boundary-value-problem assumes that at the starting point of the
magnetic field line, i.e. at minimal $\xi=\xi_{-}$, the boundary
value $a_{+0}$ should be given for $a_+$ wave, propagating in the
direction of increasing $\xi$: $a_+(\xi=\xi_-)=a_{+0}$. For the oppositely propagating wave the boundary value, $a_{-0}$, should be given at the right end point 
$\xi=\xi_{+}$, of the magnetic field line section,
$[\xi_{-},\xi_{+}]$: $a_-(\xi=\xi_+)=a_{-0}$. For the case of the closed magnetic field line, starting and ending at the solar surface, $a_{+0}=a_{-0}=1$
because of our choice of the Boundary Condition (BC) for the Poynting
flux. A few examples of the problem formulation for Eq.(\ref{eq:wkbrefl1d}) are delegated to Section 3.
 
 In Eq.(\ref{eq:energytotal1d}) one can express $(\Gamma_-w_-+\Gamma_+w_+)/B=2(a_-+a_+)a_-a_+d\xi/ds\{S_A/B\}=d(a_-^2-a_+^2)/ds\{S_A/B\}$ and  on dividing Eq.(\ref{eq:energytotal1d}) by $\{S_A/B\}$,
it can be rewritten as follows:
\begin{eqnarray}\label{eq:energytotal1d1}
 \frac{2N_ik_B}{\{S_A/B\}B(\gamma-1)}\frac{\partial T}{\partial t}&+&\frac\partial{\partial s}\left[ \left\{\frac{N_iu}{\{S_A/B\}B}\right\}\frac{2k_B\gamma T}{(\gamma-1)}-\frac{\kappa_0T^{5/2}}{\{S_A/B\}B}\frac{\partial T}{\partial s}\right]=\nonumber\\
= - \frac{N_eN_i\Lambda_R(T)}{\{S_A/B\}B} &+&\frac\partial{\partial s}\left[ a_-^2-a_+^2+\left\{\frac{N_iu}{\{S_A/B\}B}\right\}\frac{GM_\sun m_p}R\right]
\end{eqnarray}
In application to the TFLM it is convenient to denote with "+" and "-" subscripts the waves propagating 
outward and inward correspondingly and assume that the variable s along the thread equals zero at the solar surface and is positive at $R _\odot<R<R_b$. These assumptions require to re-define the BCss at the interface $R=R_b$ between the TFLM and GC models as follows:
$$
b_R\vert_{R=R_b}>0:\,\left(\frac{u}B\right)_{TFLM}=
\left(\frac{({\bf B}\cdot{\bf u})}{B^2}\right)_{GCM},\,
(a_-)_{TFLM}=(a_-)_{GCM},\,(a_+)_{GCM}=(a_+)_{TFLM}
$$
$$
b_R\vert_{R=R_b}<0:\,\left(\frac{u}B\right)_{TFLM}=-
\left(\frac{({\bf B}\cdot{\bf u})}{B^2}\right)_{GCM},\,
(a_-)_{TFLM}=(a_+)_{GCM},\,(a_-)_{GCM}=(a_+)_{TFLM}
$$
\subsubsection{BCs for temperature and density}
The temperature is governed by  Eq.(\ref{eq:energytotal1d1}) which is of the second order with respect 
to the spatial coordinate. Hence, at the interface between  TFLM and GCM both temperature and its gradient should be continuous, so that the BC for temperature within the TFLM may be taken from the GCM: $T_{TFLM}=T_{GCM}$ at $R=R_b$. Accordingly, once the TFLM equations have been solved with the given temperature at $R=R_b$ and the BC at $R=R_\odot$ as discussed below, the gradient, $\left(\frac{\partial T}{\partial s}\right)_{TFLM}$, at $R=R_b$ is known and can be used to set the radial temperature gradient within the GCM. Assuming the radial component of the temperature gradient to be dominant, one has:  $\left(\frac{\partial T}{\partial s}\right)_{GCM}\approx \left(\frac{\partial T}{\partial R}\right)_{GCM}$. Hence, the equation, $\left(\frac{\partial T}{\partial R}\right)_{GCM}=\left(\frac{\partial T}{\partial s}\right)_{TFLM}/\vert b_R\vert$, may be used to close the boundary value problem in the GCM by setting the heating flux through the interface between TFLM and GCM.  

The density at the discussed interface is controlled by the direction of $u$. If $u>0$, then $\left\{\frac{N_iu}B\right\}_{TFLM}= (N_i)_{TFLM}\left(\frac{u}B\right)_{GCM}$, otherwise $\left\{\frac{N_iu}B\right\}_{TFLM}= \left(\frac{N_iu}B\right)_{GCM}$. All we need now in order to get well-posed problem for the listed 1D governing equations is to close the model with the boundary condition at "low boundary". A good opportunity is to set the boundary conditions {\it on top of the Transition Region} (TR), which may be matched with the chromosphere via an {\it analytical model} of the transition region. This model has been presented by \cite{lion01} (see also \cite{lion09} and \cite{downs10}).
 
To use this model, we choose at each magnetic thread a short section of the length of, $L_{TR}=\int_{R_\odot}^{R_{TR}}{ds}\sim 1$ Mm to be a width of the TR along the magnetic field line. This is an important distinction from previous works, in which the temperature on top of the TR had been set, rather than the TR width.  In the steady-state version of Eq.(\ref{eq:energytotal1d1}) we keep only the terms which dominate within the TR, i.e. at high density and abrupt temperature gradients: 
\begin{equation}\label{eq:transition}
-\frac\partial{\partial s}\left(\kappa_0T^{5/2}\frac{\partial
  T}{\partial s}\right) =-N_eN_i\Lambda_R(T).
\end{equation}

On multiplying Eq.(\ref{eq:transition}) by $\kappa_0T^{5/2}(\partial
T/\partial s)$ and by integrating from the interface to chromosphere
to a given point at a temperature, $T$, one can obtain:
\begin{equation}\label{eq:transition1}
\frac12
\kappa_0^2 T^5
\left(
\frac{\partial T}{\partial s}
\right)^2\vert^{T}_{T_{ch}}=
\{N_iT\}^2
\int^{T}_{T_{ch}}{\kappa_0T_1^{1/2}\Lambda_R(T_1)dT_1}.
\end{equation}
Here the product, $\{N_iT\}$ is assumed to be constant, therefore, it is
separated from the integrand, since the temperature gradient scale within the TR is much shorter than the barometric scale: $T/(\partial T/\partial s)\ll k_BT/(GM_\odot m_p/R_\odot^2)$.  On the
left hand side of the equation the half of the heat flux squared should be taken at the temperature, $T$, with the positive sign and
at temperature $T_{ch}$ with negative sign. We can neglect
the contribution from the latter term at the lower boundary of this
transition region, if we postulate that the transition region is heated 
solely by the heat transfer from the corona and the lower boundary of
the transition region is where the heat flux from the corona turns to 
zero. For given $T_{ch}\sim (1\div2)\cdot10^4$ K and $L_{TR}$ one can solve
$\{N_iT\}$  in terms of  the radiation loss integral:
\begin{equation}\label{eq:transition2}
\{N_ik_BT\}=\frac1{L_{TR}}\int_{T_{ch}}^{T_{TR}}{\frac{\kappa_0T_1^{5/2}dT_1}
{U_{\rm heat}(T_1) }},\qquad U_{\rm heat}(T)=\sqrt{
\frac2{k_B^2}\int^{T}_{T_{ch}}{  \kappa_0(T^\prime)^{1/2}\Lambda_R(T^\prime)dT^\prime}},
\end{equation} 
which also allows us to find the heat flux into the TR from the Low Corona:
\begin{equation}\label{eq:transition3}
\kappa_0 T_{TR}^{5/2}
\left(\frac{\partial T}{\partial s}\right)_{T=T_{TR}}=\{N_ik_BT\}U_{\rm heat}(T_{TR}).
\end{equation}
Here, $U_{\rm heat}$ is a quantity with the dimension of speed, such that $2L_{TR}/((\gamma-1)U_{\rm heat}(T_{TR}))$ is a typical temperature relaxation time in the TR. 
\begin{figure}\label{fig:TR}
\includegraphics[scale=0.5]{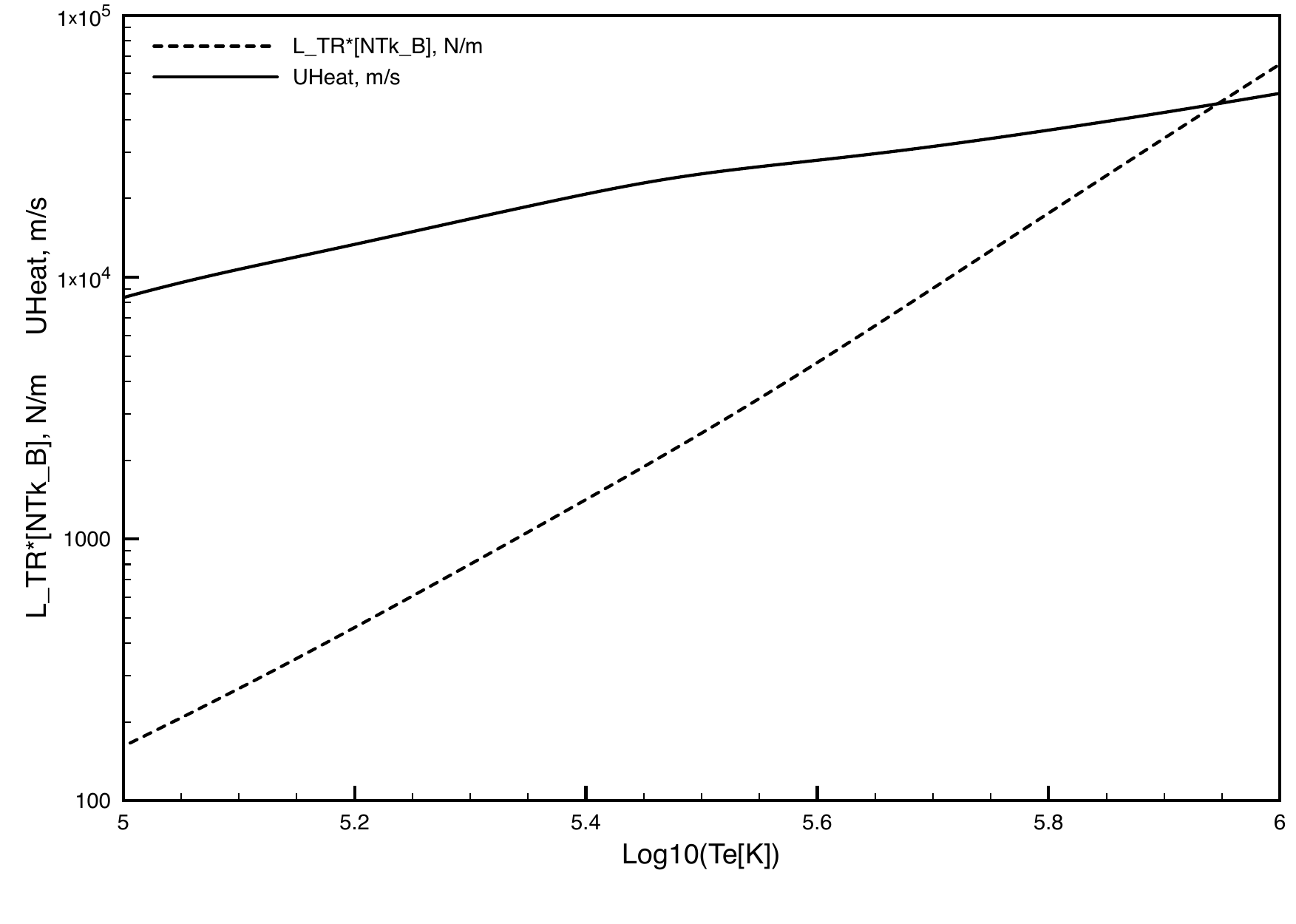}
\caption{Preprocessed CHIANTI table for radiative cooling, which allows us to formulate the boundary condition at the solar surface, for the TFLM. For a known width, $L_{TR}$, of the internal transition region and for the input temperature, $T$, the constant product, $\{NTk_B\}$ may found using the dashed curve. Then, the heat flux into the transition region, $U_{\rm heat}\{NTk_B\}$, may be found using also the solid curve.}
\end{figure}

We arrive at nonlinear boundary conditions on top of the TR, which for the known width $L_{TR}$, of the TR and for a given temperature, $T_{TR}$, on top of the TR allow us to find the heat flux and pressure there using Eqs.(\ref{eq:transition2},\ref{eq:transition3}). Note, that the BCs, which set neither temperature nor its gradient, however, relate the temperature and its gradient at the boundary, are unusual, but mathematically legitimate. These BCc may be easily implemented if the temperature functions in the right hand side of Eqs.(\ref{eq:transition2},\ref{eq:transition3}) as well as the function, $\Lambda_R(T)$, are all tabulated using the CHIANTI database \citep{landi13} (see Fig.~1). Now, the TFLM is fully described and designed to be solved numerically. There is still a minor uncertainty in the way to distinguish the TR from the top of chromosphere, originating from the fact, that we do not include a consistent chromosphere model. Particularly, 
 the TR solution at some point should be merged to the chromoshere solution with no jump in 
pressure, at the location, which depends both on $\{N_Ik_BT\}_{TR}$ and on the pressure barometric distribution in the chromosphere. However, the uncertainty in this location, which also results in some
uncertainty in $L_{TR}$, is negligible, because the barometric scale in the chromosphere is small.
\section{Analytical Solutions and Scaling Laws for the Alfv\'en Wave Turbulence in the Low
  Solar Corona}
  As has been demonstrated above, the main point of the developed approach is to achieve efficient and realistic modeling of the solar atmosphere.  On the other hand, the analytical solutions, discussed in the present Section are over-simplified and are relevant only for analyzing the equations describing the Alfv\'en wave dynamics. Although these solutions are not directly usable 
 for doing simulations, they may give hints on dependencies between the different model parameters. Of a particular interest is the question, which model parameters should be modified to achieve a better agreement with the observations of the solar wind parameters at 1 AU. Therefore, we provide here some solutions describing the wave turbulence in the different regions (closed vs open field lines, lower vs global SC etc).   
\subsection{Solution for Coronal Holes - Weak Reflection}
Here, we consider an open magnetic field line, by assuming as we did above, that the wave of amplitude, $a_+$, propagates outward the Sun. First,
consider the case when the reflection due to a gradual change in the
Alfv\'en speed magnitude (the latter is assumed to exponentially decay
outward the Sun, $V_A=V_{A0}\exp(-s/L_{V_A})$) is small compared with
the characteristic dissipation rate:
$\frac12\frac{ds}{d\xi}\frac{|{\cal R}|}{V_A}=\frac12\frac{d\log V_A}{d\xi}\ll
a_+$. This assumption is valid at sufficiently high altitudes much above the transition region (where, to the contrary the abrupt density gradients cause large reflection as we discuss below). As the result of a weak reflection, the amplitude of the wave reflected back and propagating toward the Sun is small:
 $a_-\ll a_+$. The governing equations in this limiting case read:
\begin{equation}\label{eq:weakrefl}
\frac{da_+}{d\xi}=\frac12\frac{d\log V_A}{d\xi}a_--a_-a_+,\qquad \frac{da_-}{d\xi}=\frac12\frac{d\log V_A}{d\xi}a_++a_-a_+
\end{equation}
For the exponentially decaying profile of the Alfv\'en speed, by
introducing a constant small parameter:
$$
C_{\rm refl}=-\frac12\frac{d\log
  V_A}{d\xi}\sqrt{\frac{V_{A0}}{V_A}}=\frac12
\sqrt{\frac{\{L_\perp\sqrt{B}\}^2V_{A0}}{L_{V_A}^2\left\{{S_A}/B\right\}\mu_0}}=
0.09\frac{\{L_\perp\sqrt{B}\}}{150\,{\rm
      km\,T^{1/2}}}\frac{R_\odot}{L_{V_A}}
\sqrt{\frac{V_{A0}}{10^3{\rm
          km/s}}\frac{1.1\cdot10^6\frac{\rm W}{\rm m^2T}}{\left\{S_A/{B}\right\}}},
$$
one can easily find the solution of Eqs.(\ref{eq:weakrefl}), tending to zero at infinity:
$$a_+=\sqrt{\frac{V_A}{V_{A0}}},\qquad a_-=\frac{C_{\rm refl}}{1+C_{\rm
    refl}}a_+\approx C_{\rm refl}a_+.$$
We found that, for the exponential profile of the Alfv\'en speed, 
the small ratio of the amplitude of the incoming wave to that for the
outgoing wave is constant. This observation, in principle,  allows us to calculate
the dissipation rate for the dominant wave without calculating the
amplitude of the reflected wave, since 
$\Gamma_+=\frac2{L_\perp}\sqrt{\frac{w_-}\rho}\approx C_{\rm refl}\frac2{L_\perp}\sqrt{\frac{w_+}\rho}$.  
The latter comment, although valid only for a particular case of exponentially decaying $V_A$, allows
us to link the current model to that  described in \cite{sok13}, where we also parameterized the turbulence dissipation
within the coronal holes using small dimensionless $C_{\rm refl}$. The WKB 
approximation we used in that paper predicted no inward propagating waves originating from the open magnetic field lines ($w_-=0$). Therefore, we assumed therein a small but finite (due to reflection) amplitude of the inward propagating wave to be parameterized as $w_-=C_{\rm refl}^2w_+$, so that:  
\begin{equation}
\frac{\partial w_\pm}{\partial t}+\nabla\cdot[({\bf u}+{\bf V}_A)w_\pm]+\frac{w_\pm}2(\nabla\cdot{\bf u})
=-\frac2{L_\perp}\sqrt{\frac{\max(w_\mp,C^2_{\rm refl}w_\pm)}\rho}w_\pm.
\end{equation}
In \cite{sok13} we did not discuss the reflection mechanism, so that $C_{\rm refl}$ was an arbitrary and uncertain free
parameter. In the model developed here we calculate the reflection
coefficient ${\cal R}$  for realistic
distribution of the magnetic field and plasma parameters, to greatly
reduce the model uncertainty, however, we see that the choice of
$C_{\rm refl}={\rm const}\approx0.01\div0.1$ in \cite{sok13} was reasonable and might be
derived analytically. 

In a general case of an arbitrary (not necessarily exponential) profile of the Alfv\'en wave speed, Eqs.(\ref{eq:weakrefl}) still can be solved at small (not necessarily constant) value of $C_{\rm refl}$. Indeed, the total of two Eqs.(\ref{eq:weakrefl}) is a linear and easy-to-integrate equation, which gives: $(a_++a_-)\propto\sqrt{V_A}$, so that $a_+\approx\sqrt{V_A/V_{A0}}$ as long as 
$a_-\ll a_+$, with constant $V_{A0}$, for a given thread being a characteristic value of the Alfv\'en wave speed at low altitude.    Then, in the second of Eqs.(\ref{eq:weakrefl}) the left hand side is quadratic in small $C_{\rm refl}$, as the small ($\propto C_{\rm refl}$) derivative of a smaller amplitude, $a_-\sim C_{\rm refl} a_+$. Therefore, the two linear in $C_{\rm refl}$ terms on the right hand side should cancel each other, which requirement gives: $a_-\approx -\frac12\frac{d\log V_A}{d\xi}$. In this way, we arrive at a simple and transparent estimate for the wave energy density within the coronal holes, which linearly scales with the magnetic field:
$w_+=\left\{S_A/{B}\right\}\sqrt{\mu_0\rho}a^2_+=\left\{S_A/{B}\right\}B/{V_{A0}}\propto B$. From Eq.(\ref{eq:heatingf}), we can now derive
the following expressions for the heating function: $(\Gamma_+w_++\Gamma_-w_-)\approx a^2_+a_-\left(2\left\{S_A/B\right\}B\sqrt{\frac{\left\{S_A/B\right\}\mu_0}
{\{L_\perp\sqrt{B}\}^2V_A} }\right) \approx\left\{S_A/{B}\right\}B\frac{V_A}{V_{A0}}\left|\frac{d\log V_A}{ds}\right| ={\cal R}w_+\propto B^2/(L_{V_A}\sqrt{\rho})$.   

Note also that for the Alfv\'en speed profile gradually increasing in the
outward direction, the solution for the dominant wave energy density
is: $w_+=\left\{S_A/{B}\right\}\sqrt{\mu_0\rho}V_{A0}/V_A\propto \rho/B$. The heating function is: $(\Gamma_+w_++\Gamma_-w_-)\approx\left\{S_A/{B}\right\}B\frac{V_{A0}}{V_{A}}\left|\frac{d\log V_A}{ds}\right| ={\cal R}w_+\propto \sqrt{\rho}/L_{V_A}$. 

Now, we arrive at an important conclusion. Within the coronal holes both the distribution of the turbulence energy and the heating function do not depend on the dissipation length, $L_\perp$, as long as the minor wave amplitude  and the dissipation rate are fully controlled by the wave reflection. The latter,  in turn, is fully controlled by the field and plasma profile. The model of a self-consistent plasma state within the coronal hole, which is controlled by the heating function dependent only on the plasma state itself, seems to be reasonable and physics-based. 

\subsection{Coronal Holes: Strong Reflection}
\begin{figure}\label{fig:1}
\includegraphics[scale=0.5]{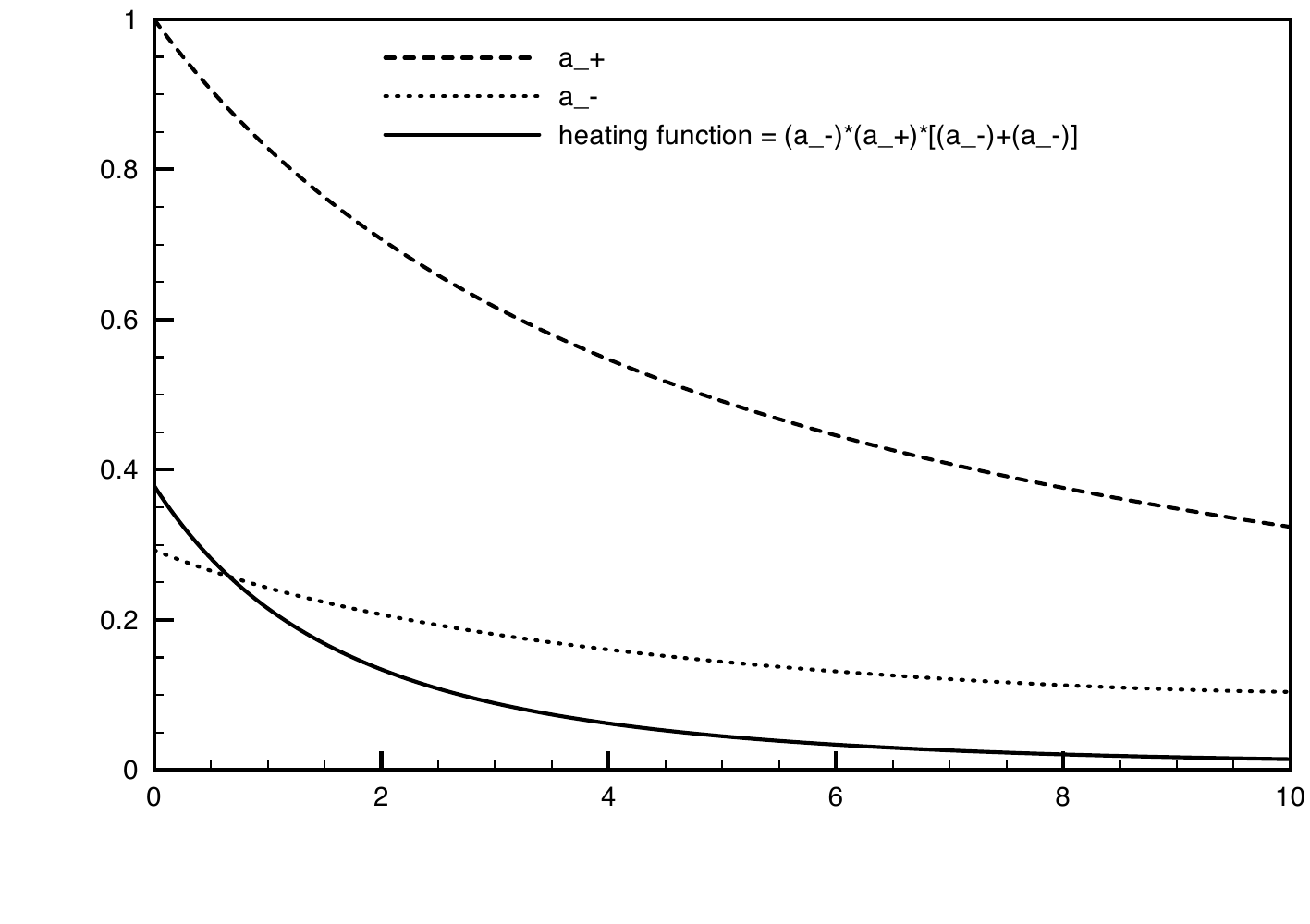}
\caption{Dimensionless amplitudes of the Alfv\'en wave
  turbulence in a coronal hole with a strong wave reflection,
$a_\pm=\sqrt{w_\pm/\left(\left\{S_A/{B}\right\}\sqrt{\mu_0\rho}\right)}$,
(dashed/dotted lines) and the dimensionless heating function,
  $\sum{\Gamma_\pm w_\pm}/\left(2\left\{S_A/B\right\}B\sqrt{\frac{\left\{S_A/B\right\}\mu_0}
{\{L_\perp\sqrt{B}\}^2V_A} }\right)$ (solid line)
  are presented as functions of the
  effective magnetic field line length, $\xi(s)=\int_0^s{
\sqrt{\frac{\{S_A/B\}\mu_0}{\{L_\perp\sqrt{B}\}^2V_A(s^\prime)}}ds^\prime}$.}
\end{figure}

In the case of strong reflection, in which case the reflection
coefficient in Eq.(\ref{eq:reflection}) is bounded with the cascade rate the
governing equations for the wave dimensionless amplitude read:
$$
\frac{da_+}{d\xi}=-a_-(a_+-2a_-)-a_-a_+,\qquad \frac{da_-}{d\xi}=-a_+(a_+-2a_-)+a_-a_+
$$
Their exact analytical solution for the coronal hole, which should tend to zero at
$\xi\rightarrow\infty$ has a constant amplitude ratio, $a_-/a_+=q<1/2$. This ratio can be
easily found by requiring that $d(a_-/a_+)/d\xi=0$, so that
 $q=1-\sqrt2/2\approx0.29$,
$a_+=1/[1+2q(1-q)\xi]\approx1/(1+0.42\xi)$,
$a_-=q/[1+2q(1-q)\xi]\approx0.29/(1+0.42\xi)$. These functions and the
product, $a_-a_+(a_-+a_+)$ representing the heating function, $\sum{\Gamma_\pm w_\pm}/\left[2\frac{B^{1/2}\rho^{1/4}\mu_0^{3/4}}
{\left\{L_\perp\sqrt{B}\right\}}\left\{S_A/{B}\right\}^{3/2}\right]$, are plotted
in Fig.
%~\ref{fig:1}
2.  
The heating function is maximal near the solar surface and decays with the heliocentric distance as $\propto B^{1/2}\rho^{1/4}/(1+0.42\xi)^3$.  
\subsection{Scaling Laws for the TFLM}
The weakness of any model relying on the solar magnetogram is uncertainty of the solar magnetic field observations. We introduce the parameter of the model, $\{S_A/B\}\approx 1.1\cdot10^6{\rm W/(m^2T)}$ with one digit after a decimal period - is this legitimate? How accurate is the observed magnitude of the solar magnetic field to be multiplied by this model parameter? Numerous observatories provide different values for the measured field. The region in the solar wind which is determined by the polar coronal holes may be large and any realistic model of the solar wind should account these holes, however, the solar magnetic field measured in these holes, by many reasons, may be unrealistically low.

To mitigate the effect of too low and, probably, underestimated magnetogram field, we apply some scaling factor $B_{\rm scale}\ge1$ in our simulations, so that the observed solar magnetic field multiplied by $B_{\rm scale}$ (which is $\approx3$ for GONG magnetograms) is used as the boundary condition of the model: $B_{\rm TFLM}|_{R=R_\odot}=B_{\rm scale}\cdot B_{\rm observed}|_{R=R_\odot}$. We note that the TFLM equations are not affected by this scaling if in accordance with 
increasing the magnetic field we also decrease the model parameters:
\begin{equation}
B\rightarrow B\cdot B_{\rm scale},\qquad \{S_A/B\}\rightarrow\{S_A/B\}/B_{\rm scale}, \qquad \{L_\perp \sqrt{B}\}\rightarrow\{L_\perp \sqrt{B}\}/B_{\rm scale},
\end{equation}   
because the magnetic field in the TFLM equations is present only in combinations, $\{S_A/B\}B$ and $\{S_A/B\}/(B\{L_\perp\sqrt{B}\}^2)$.
\section{Simulation Results and Discussion}
All simulations were performed with the Space Weather Modeling Framework (SWMF - see {\color{cyan} {\it T\'oth et al.} (2004, 2005, 2007, 2012)}. The SWMF included the models (components) to simulate the Solar Corona and Inner Heliosphere, both models accounting for the contributions from the Alfv\'en wave turbulence as we described in \cite{sok13}, \cite{oran13} and \cite{vanderholst13} (the AWSoM model). The most important distinction in the current simulations is that we apply the AWSoM model only to the GCM, while the transition region and lower corona are described using the TFLM. 

\begin{figure}[ht!]
  \centering
  \includegraphics[height=4.5in]{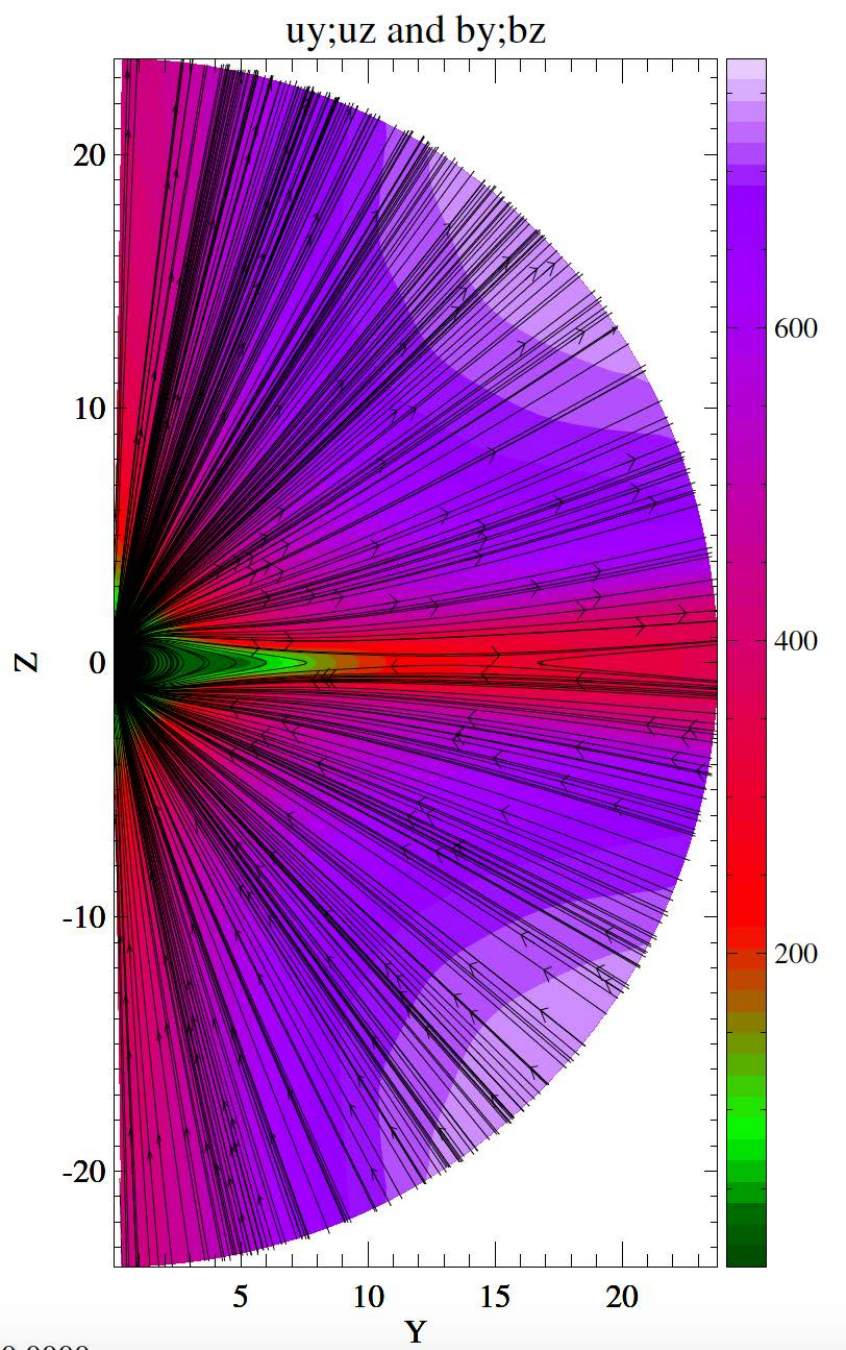}
  \includegraphics[height=4.5in]{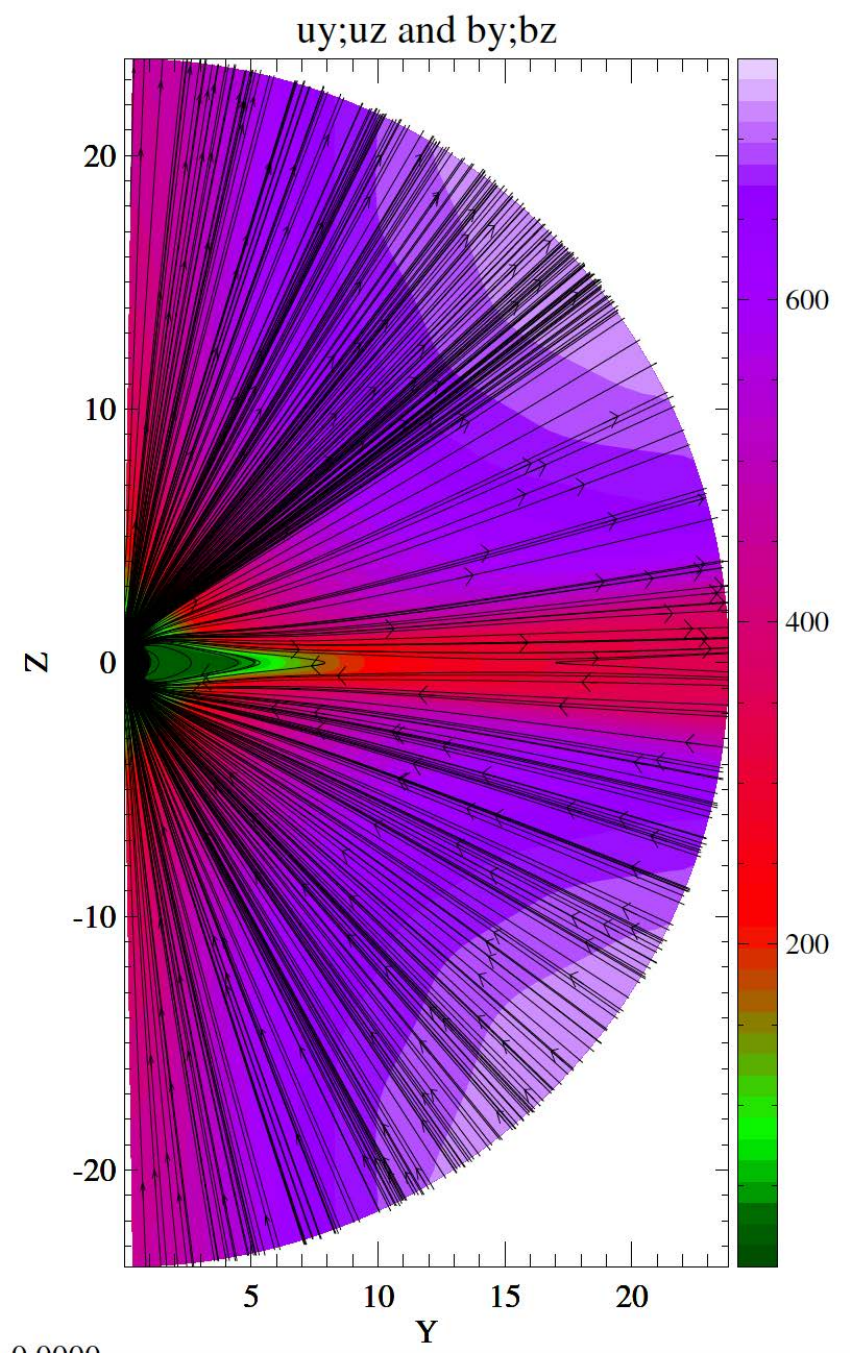}
  \caption{Steady-state solar corona with the dipole field, after 18000 iterations.  Solar wind speed is shown with a color scale, 
  black lines are the magnetic field lines.  
  Left panel: simulation with the AWSoM model with the inner boundary at $R=R_\odot$. Right panel: simulation with TFLM+GCM, the interface to GCM is at $R=1.1R_\odot$. The results in the GCM are practically identical with those in the AWSoM.}
  \label{fig:comparison}
\end{figure}

In this way we benefit from saving the computational resources which otherwise should be spent to resolve the true structure of
the transition region using a highly refined grid. We start from the observation that the gain in computational efficiency is achieved with no degrade in the accuracy and quality of the numerical results. In Fig. \ref{fig:comparison} we show a comparison of the numerical results obtained with the ASWoM model (see  \cite{sok13}, \cite{oran13} and \cite{vanderholst13}) and presented in the left panel, with the new result obtained with the TFLM (the AWSoM-R model). The grid for the  AWSoM run requires much finer grid cells to resolve the transition region. In addition, in time-dependent runs these finest grid cells control and severely reduce the time step making it as short as a few ms. At the same time, in the AWSoM-R result (the right panel) there is no noticeable difference from that obtained with AWSoM, except that it may be obtained much faster and with the time step about one second, which is equivalent to a gain by a factor of several hundreds in the computational efficiency. The capability to do simulations significantly faster  than the real time on a couple 100 CPU cores  is the main advantage of the  AWSoM-R with TFLM.      

\begin{figure}[ht!]
  \centering
  \includegraphics[scale=0.5]{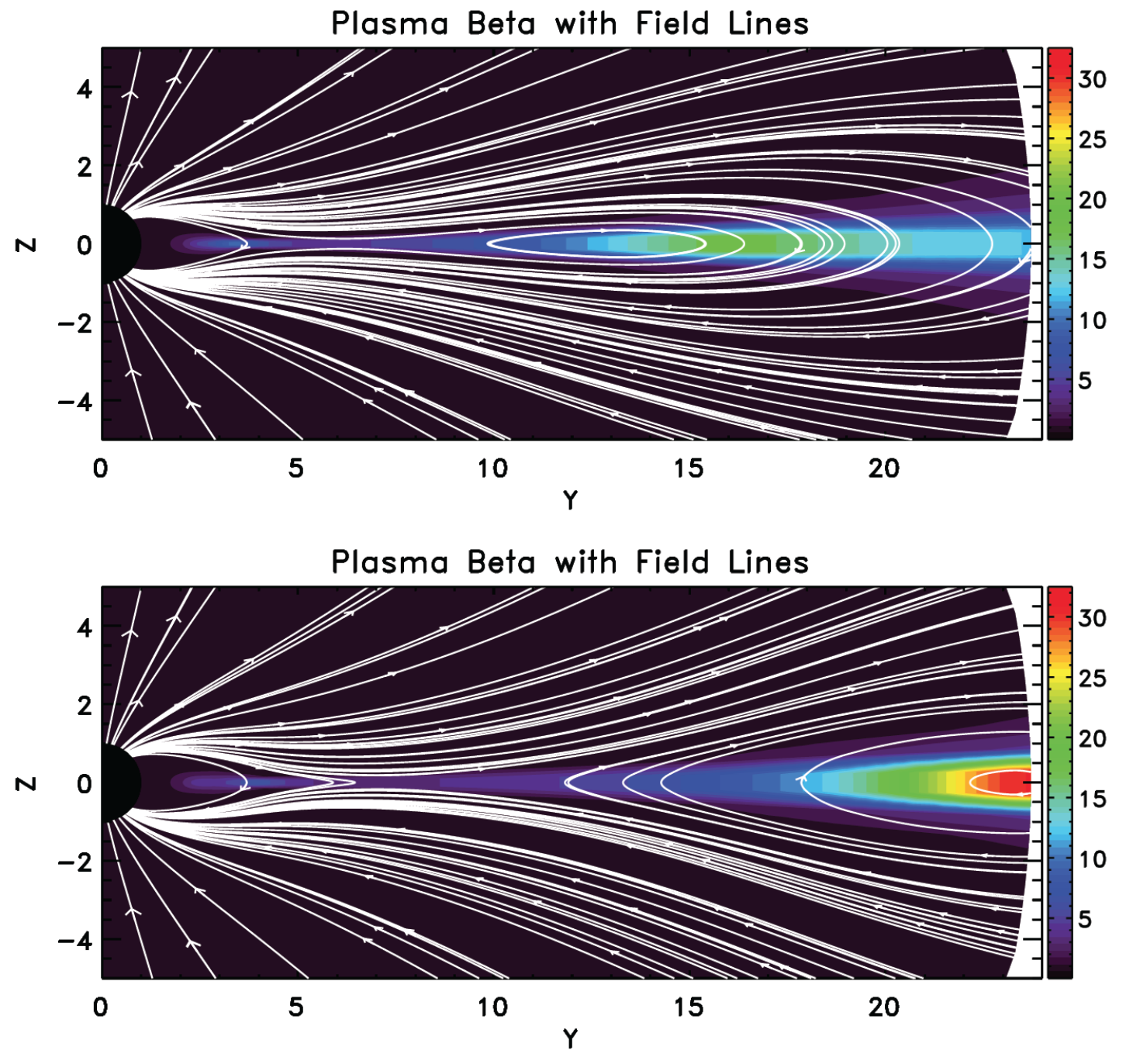}
  \caption{Variation of plasma beta and magnetic field lines along XY plane in HGR coordinates taken 6 hours 40 minutes apart. As plasma pressure increases, it stretches out the field lines (upper panel) and the reconnected field lines move anti- Sunward (lower panel)}
  \label{fig:blob}
\end{figure}

The evolution of the solar magnetic field is comparatively slow, therefore, it is usually believed that the steady-state solution with the time independent boundary condition such as those shown in Figure 3 should be a good approximation (see, for example, steady-state solutions in \cite{oran13}). However, it appears that, depending on the resolution in different parts of the computational domain, a good convergence in the places of interest may or may not be achieved for  the numerical steady-state solution at moderate number of iterations (less than $\approx 20,000$ for the case presented in Figure 3): note a reduced solar wind speed at the pole, where due to small grid size the convergence is slowest. With the larger number of iterations, the magnetic reconnection occurs near the top of helmet streamer (at $y\approx5R_\sun$) occurs,  thus preventing us to achieve a better steady-state solution. The advantage of time-dependent solution as we advocate here is in the capability to describe the time-dependent processes that occur even with the steady-state boundary condition for the magnetic field. 

To demonstrate this capability we simulated 10 days of evolution in the Solar Corona with the steady-state dipole magnetic filed. Such simulation reveals a dynamic helmet streamer structure, which periodically produces plasmoids known as "streamer blobs" \citep{wang00}. Our investigation shows that these blobs form as a result of pressure imbalance mainly because of increased ion temperatures at the streamer top. (see Fig.\ref{fig:blob}). 

\begin{figure}[ht!]
  \centering
  \includegraphics[scale=0.5]{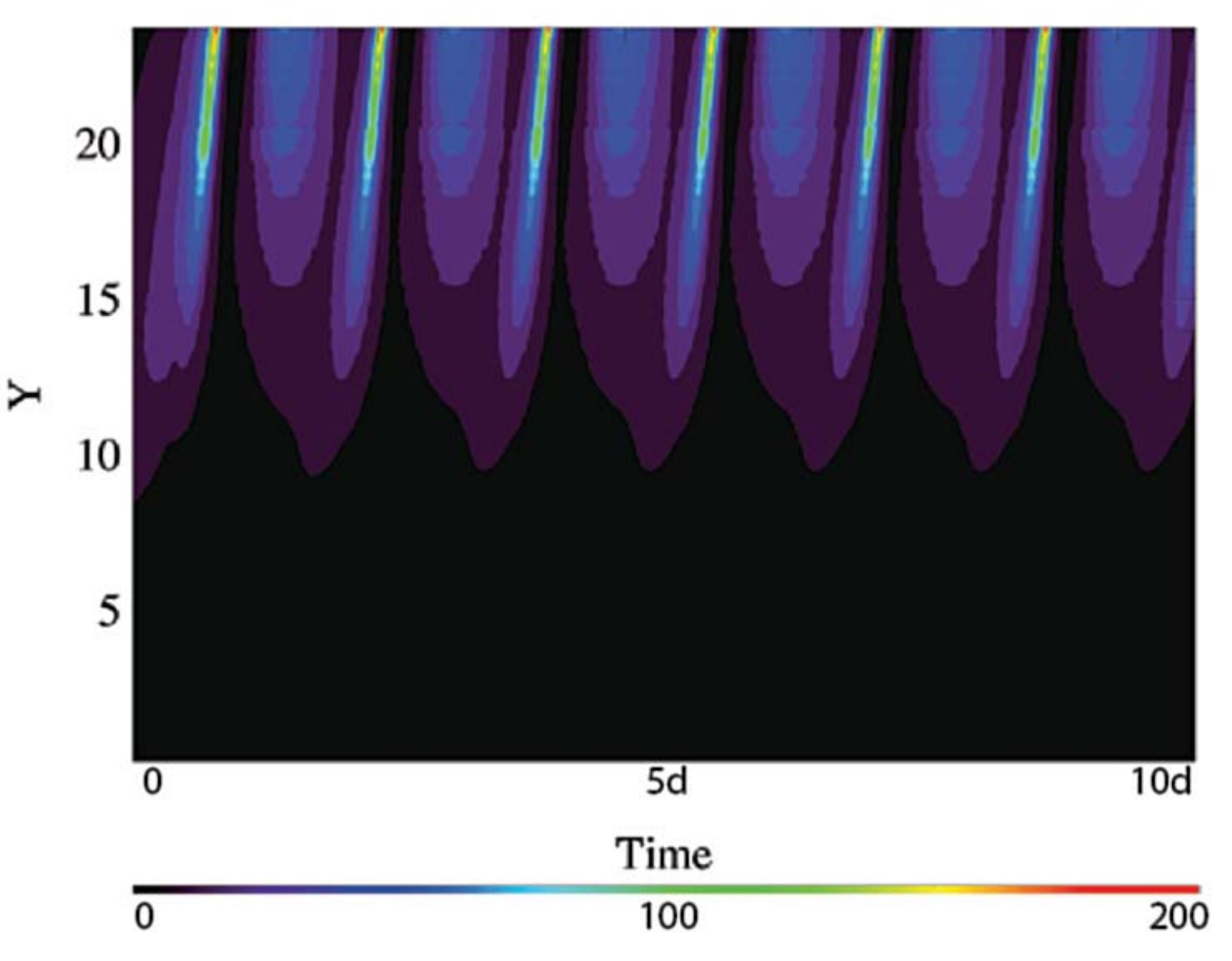}
  \caption{Plasma beta variation in the Solar Corona from the inner boundary to 25 Rs along the y-axis of the HGR coordinate system obtained from the simple dipole simulation for 10 days of physical time. }
  \label{fig:6blobs}
\end{figure}

To find a typical time of the blob formation, we choose a radial line in the equatorial plane rotating with the Sun (as an example we used y-axis of the HelioGraphic Rotating (HGR) coordinate system, in which longitude is defined as the Carrington longitude) and visualize the distribution of plasma beta along this line as a function of time (x-axis) and radial coordinate(y-axis) as shown in Fig.\ref{fig:6blobs}. The plasma beta starts increasing at the heliocentric distance of $\approx$9$R_s$ implying the start of the disconnection event and the disconnected blob moves anti-Sunward. These intermittent detachment events occur with a periodicity of about 40 hours (six times within 10 days), which is in a good agreement with the observations.  We, thus, show that  in the self-consistent \alf Wave Turbulence based  model the slow solar wind is intermittent even if the solar magnetic field is steady-state and perfectly symmetric. 
\begin{figure}[ht!]
  \centering
  \includegraphics[width=0.5\textwidth]{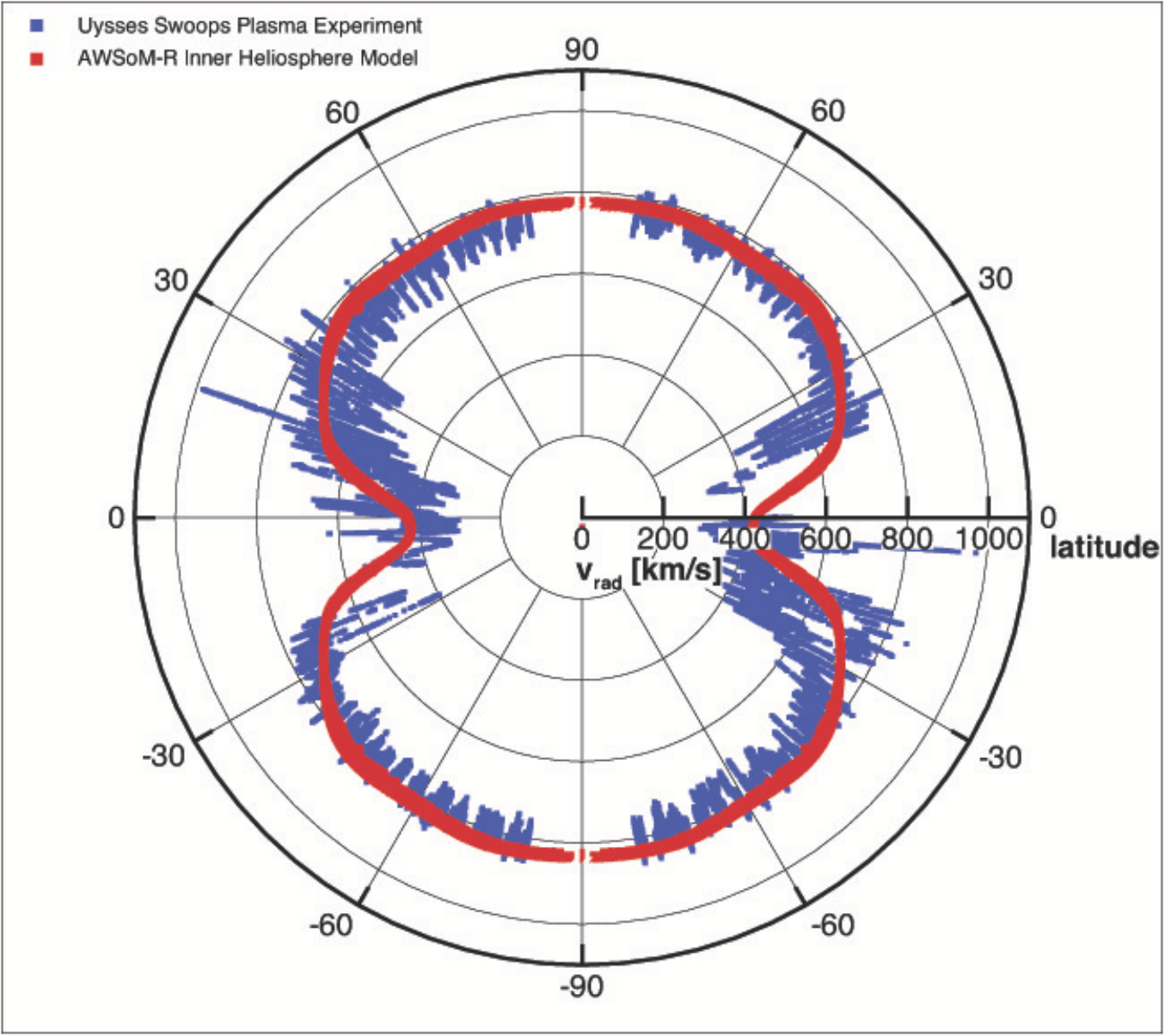}
  \caption{Polar scatterplot of the radial velocity (radius) versus %the
    latitudinal location of the spacecraft Ulysses (angle). Blue denotes
    Ulysses SWOOPS data from between 1990 and 1997. Red shows the model result;
    the data points were sampled at the same radial distances as
    Ulysses passing during this period.}
  \label{fig:ulysses}
\end{figure}

We did a much longer simulation to test the AWSoM-R model for the inner heliosphere. Specifically, 
using a simple dipole magnetic field as the boundary condition for the radial
magnetic field component, we simulated over 47 days of
physical time with the AWSoM-R model.
We sampled plasma parameters at the same distances
where the Ulysses spacecraft had passed. Figure~\ref{fig:ulysses} shows the
comparison of radial velocity distribution along different latitudes between
the SWOOPS \citep{Bame:1992} plasma measurements (blue) and the simulation
results (red). We selected observations of the time frame between the years 1990
and 1997 so that we cover the complete range of latitudes. The observations
are during the declining phase of the solar cycle. The solar wind
distribution shows the clear difference between slow and fast solar wind
regions. Both these wind speeds as well as the transition latitudes are
captured by the AWSoM-R model. 

\section{Conclusion}
The AWSoM-R model presented here extends the earlier developed AWSoM (\cite{sok13} and \cite{vanderholst13}) with the TFLM description for the transition region and Low Solar Corona. It allows us to simulate the Solar-Earth environments on realistic 3D grids faster than real time and with no loss in the results quality.

\section{Acknowledgment}
The collaboration between the CCMC and University of Michigan was supported by the NSF SHINE grant 1257519 (PI Aleksandre Taktakishvili). The work performed at the University of Michigan was partially supported by National Science Foundation grants AGS-1322543 and PHY-1513379, NASA grant NNX13AG25G, the European Union's Horizon 2020 research and innovation program under grant agreement No 637302 PROGRESS. This work was also supported by a NASA Heliophysics DRIVE Science Center (SOLSTICE) at the University of Michigan under grant NASA 80NSSC20K0600. We would also like to acknowledge high-performance computing support from: (1) Cheyenne  sponsored by the National Science Foundation, and (2) Pleiades operated by NASA?s Advanced Supercomputing Division.
\clearpage

\end{document}